\documentclass[paper,12pt]{article}

\usepackage[pdftex]{graphicx}
\pdfoutput=1
\usepackage{amsmath}
\usepackage{mathtools}
\usepackage{braket}

\usepackage{tikz}
\usepackage{cancel}
\usepackage{amsfonts}
\numberwithin{equation}{section}

\begin{document}

\setlength{\baselineskip}{0.7cm}
\begin{titlepage}
    \begin{flushright}
        NITEP 167
        
    \end{flushright}
    \vspace{10mm}
    \begin{center}
        \Large\textbf{Wilson-line Scalar Mass } \\
       \vspace{2mm}
        \Large\textbf{in Flux Compactification on an Orbifold $T^2/Z_2$} \\
    \end{center}
    \vspace{10mm}
    \begin{center}
        {\large Nobuhito Maru},$^{a,\,b}$
        and {\large Hiroki Tanaka}$^{\,a}$
    \end{center}
    \vspace{0.2cm}
    \begin{center}
        ${}^{a}$ \textit{Department of Physics, 
        Osaka Metropolitan University, \\
        Osaka 558-8585, Japan} \\
        ${}^{b}$ \textit{Nambu Yoichiro Institute of Theoretical and Experimental Physics (NITEP), \\
        Osaka Metropolitan University, Osaka 558-8585, Japan} \\
    \end{center}
    \date{}
\vspace*{5mm}
    \begin{abstract}
In a torus compactification with magnetic flux, 
 the scalar field generated from the higher dimensional gauge field becomes 
 the Nambu-Goldstone  boson of the translational symmetry in extra spaces and 
 the mass of the scalar field is not allowed. 
In this paper, we explicitly show that the one-loop quantum correction to the mass 
 of the scalar field is generated at fixed points
 in a case of an orbifold compactification $T^2/Z_2$ with magnetic flux. 
This is because the scalar field is a pseudo Nambu-Goldstone boson 
 as a result of explicit breaking of the translational symmetry by the fixed points. 
This result might be possible to shed a new light on the solution of the hierarchy problem.     
 \end{abstract}
\end{titlepage}

\setcounter{tocdepth}{3}
\clearpage

\section{Introduction}
Flux compactification is one of the attractive scenarios beyond the Standard Model (SM) 
 since the chiral fermion is realized, the number of generations of quarks and leptons might be predicted, 
 and so on. 
On the other hand, 
 it is also interesting to apply this scenario to the hierarchy problem in the SM.  
In this scenario, the zero mode of extra component of the higher dimensional gauge field, 
 which we call as Wilson-line (WL) scalar, is regarded as the SM Higgs field. 
However, the WL scalar mass at tree level and the quantum level vanishes 
 in a theory compactified on $T^2$ with magnetic flux 
 since the WL scalar can be considered to be Nambu-Goldstone (NG) boson of the translational symmetry 
 in extra spaces $T^2$ with magnetic flux. 
This property was shown at one-loop in a six dimensional QED compactified on $T^2$ with magnetic flux 
 in supersymmetric or non-supersymmetric cases \cite{BDDS, BDD}. 
This cancellation was investigated carefully in \cite{GL}. 
The non-Abelian extension of this property has been achieved in \cite{HM1, HM2}. 
Two-loop analysis has been carried out and the result has been shown to vanish \cite{HS}.

In order to be available to the hierarchy problem, 
 the nonvanishing mass of the WL scalar field has to be generated 
 by breaking the translational symmetry in extra spaces explicitly. 
As such attempts, the finite WL scalar mass was obtained in a six dimensional scalar QED compactified 
 on $T^2$ with magnetic flux 
 by classifying the interactions which break the translational symmetry explicitly and provide the finite WL scalar mass \cite{HM3}. 
Alternatively, the nonvanishing scalar potential of the WL scalar field was obtained and 
the pattern of gauge symmetry breaking was studied
 in a six dimensional Yang-Mills theory compactified on $T^2$ 
 by considering the magnetic flux and the constant vacuum expectation value (VEV) simultaneously, 
 which also breaks the translational symmetry explicitly.  

In this paper, a further alternative way of explicit breaking will be employed. 
We consider an orbifold compactification with magnetic flux, 
 where there exist some fixed points in extra spaces under the discrete symmetry of the orbifold. 
The translational symmetry in extra spaces is explicitly broken at fixed points, in other words, 
 the compactification itself breaks the symmetry explicitly. 
Therefore, it is expected from this observation that the nonvanishing WL scalar mass is generated at fixed points 
only from the zero mode contributions. 
We will show that this is indeed the case in a simple model.

\section{Setup}

We consider a six dimensional scalar QED compactified on an orbifold $T^2/Z_2$ with magnetic flux 
 to understand the essential points of the mechanism and avoid an unnecessary technical complexity. 
The action is given by 
\begin{align}
S &=\int d^6x \left(
-\frac{1}{4} F^{MN} F_{MN} 
-(D^M \Phi)^* D_M \Phi
\right) \\
&= \int d^6x \left[
\left(-\frac{1}{4} \right) 
\left(
F^{\mu\nu} F_{\mu\nu} +2(F^{\mu5} F_{\mu5} + F^{\mu6} F_{\mu6} + F^{56} F_{56}) 
\right) \notag \right. \\
& \left.  \qquad -(D^\mu \Phi)^* (D_\mu \Phi) -(D^m \Phi)^* (D_m \Phi)
\right],
\label{scalarQED}
\end{align}
where $M,N=0,1,2,3,5,6$ are indices in six dimensional spacetime,  
 $\mu, \nu =0,1,2,3$ are indices in four dimensional spacetime and $m=5,6$ is an index of extra spaces.   
The metric convention is taken as follows $\eta_{MN} = {\rm diag}(-1,+1,+1,+1,+1,+1)$.
The first term is the gauge kinetic term and its field strength of gauge field $A_M$ is 
\begin{align}
F_{MN} = \partial_M A_N - \partial_N A_M. 
\end{align}
$\Phi$ is a complex scalar field in the six dimensional bulk 
and their covariant derivative is 
\begin{align}
D_M \Phi = (\partial_M - ig_6 A_M)\Phi,
\end{align}
where $g_6$ is a gauge coupling constant in six dimensions with mass dimension $-1$. 
The mass of the scalar field $\Phi$ in a six dimensional spacetime is assumed to be zero for simplicity. 

The $T^2/Z_2$ orbifold flux compactification we employ in this paper is explained. 
First, $T^2$ compactification is obtained 
by imposing the periodic boundary conditions in $x_{5,6}$ direction 
as $x_{5,6} = x_{5,6} + 2\pi R$. 
Here, we take a common radius $R$ of $x_{5,6}$ directions, for simplicity. 
$Z_2$ identification is carried out as follows. 
$Z_2$ parity transformation is given by
\begin{align}
\begin{split}
x_5\rightarrow -x_5, \quad
x_6\rightarrow -x_6
\end{split}
\end{align}
and the action is invariant under these transformations. 
Combining the periodic boundary conditions and $Z_2$ parity introduced above, 
we find four fixed points in compactified space as 
\begin{align}
(x_5,x_6)=(0,0), (\pi R,\pi R), (0,\pi R), (\pi R,0). 
\end{align}

The periodic boundary conditions and $Z_2$ parity assignments have to be imposed on the fields introduced in our model. 
For the gauge field, the following conditions are taken. 
\begin{align}
A_M(x_\mu, x_5+2\pi R, x_6 + 2 \pi R) &=  A_M(x_\mu, x_5, x_6), 
\end{align}
and
\begin{align}
A_5(-x_5, -x_6)&=A_5(x_5,x_6),\\
A_6(-x_5, -x_6)&=A_6(x_5,x_6),\\
A_\mu(-x_5, -x_6)&=-A_\mu(x_5,x_6).
\end{align}
Here we note that $Z_2$ parity of $A_{5,6}$ is taken to be even by hand 
 since we are interested in the quantum corrections to the mass of their zero modes, 
 which must be even under $Z_2$ parity.  
Once the parity of $A_{5,6}$ is fixed, 
 the $Z_2$ parity of $A_\mu$ is fixed to be odd from the $Z_2$ invariance of the theory. 

The magnetic flux can be introduced by developing the VEV for $A_{5,6}$, 
 which is consistent with $Z_2$ parity. 
\begin{align}
    \braket{A_5}=-\frac{1}{2}f\epsilon(x_6) x_6 ,\qquad \braket{A_6}=\frac{1}{2} f \epsilon(x_5) x_5,  
\end{align}
where $f$ is a real constant and $\epsilon(x)$ is a sign function. 
In fact, the magnetic field in extra space is obtained 
\begin{align}
\braket{F_{56}} &= \partial_5 \braket{A_6} - \partial_6 \braket{A_5} \notag\\
&= f + f(\delta(x_5) x_5 + \delta(x_6) x_6) = f. 
\end{align}
In the last expression, we implicitly take into account 
\begin{align}
\int dx_5dx_6 \delta(x_5)x_5=\int dx_5dx_6 \delta(x_6)x_6=0. 
\label{deleps}
\end{align}
Furthermore, the magnetic flux is quantized as in the two dimensional quantum mechanics in a magnetic field. 
\begin{align}
    \frac{q}{2\pi }\int_{T^2}F=\frac{q}{2\pi }\int_{T^2}dx_5dx_6\braket{F_{56}}
    =\frac{2L^2q}{\pi }f =N \in \mathbb{Z}, 
\end{align}
where $L=\pi R$ and $q$ is an electric charge. 
We note that the integer $N$ represents the degeneracy of each mode of the field  
 since the energy spectrum is independent of $N$. 
In application to quarks and leptons, this degeneracy is interpreted as the number of generations. 

In flux compactification, the algebraic calculations can be proceeded 
 analogy to the two dimensional quantum mechanics in a magnetic field. 
In particular, the creation and annihilation operators are given by the covariant derivatives, 
\begin{align}
    [a,a^\dagger] &=\frac{1}{2g_6 f}\left[i \braket{D_5} -\braket{D_6} , i\braket{D_5}+\braket{D_6} \right]\notag\\
       &=\frac{1}{2g_6f}\Bigl\{2g_6f[\delta(x_5) x_5+\delta(x_6)x_6]
       +g_6f[\epsilon(x_5)+\epsilon(x_6)]\Bigl\}\notag\\
   &=\frac{1}{2}(\epsilon(x_5)+\epsilon(x_6))
   \label{aadagger}
\end{align}
In the last expression, we took into account again the properties \eqref{deleps}. 
Hereafter, we proceed calculations in a fundamental region of $T^2/Z_2$ with $x_5>0, x_6>0$ 
without loss of generality. 
Then, the commutation relation of the creation and annihilation operators (\ref{aadagger}) 
reduces to that of the harmonic oscillator. 

The WL scalar field can be expanded around the flux background 
\begin{align}
    \phi=\frac{1}{\sqrt{2}}(A_6+iA_5)    
    =\frac{f}{\sqrt{2}}\bar{z}+\frac{1}{L}\varphi, \qquad 
      z=\frac{1}{2}(x_5+ix_6),
     \label{eq:11} 
\end{align}
and using the creation and annihilation operators, 
the action of the bulk scalar field relevant for later calculations can be expressed as follows. 
\begin{align}
    S_{{\rm scalar}} &\supset \int d^6x 
    \Bigr(
    -(D_\mu \Phi)^* D^\mu \Phi - \bar{\Phi} H_2 \Phi \notag\\
    &\qquad\qquad
    -\frac{\sqrt{2}ig_6}{L}\sqrt{2g_6f}\overline{\varphi}\bar{\Phi}a^\dagger\Phi 
    + \frac{\sqrt{2}ig_6}{L}\sqrt{2g_6f}\varphi\bar{\Phi}a\Phi
    -\frac{2g_6^2}{L^2}\overline{\varphi}\varphi\bar{\Phi}\Phi\Bigr)
    \label{scalaraction}
\end{align}
where $\varphi$ is a 
 WL scalar field. 
$H_2 = -\braket{D_5}^2 -\braket{D_6}^2$ is a Kaluza-Klein (KK) mass operator of the bulk scalar field. 

We emphasize here that the local mass term of the WL scalar field 
 is forbidden by the gauge symmetry since it is not invariant under the gauge transformation 
 $\phi \to \phi + \partial \alpha$ 
 where $\alpha(x_\mu, x_5, x_6)$ is a transformation parameter. 
Note that this property can be also applied to the local mass terms on the fixed points 
 since the gauge transformation $\phi \to \phi + \partial \alpha_{odd}$ 
 by the $Z_2$ odd transformation parameter is operative on the fixed point 
 as a remnant of the higher dimensional bulk gauge symmetry \cite{GIQ}.

Furthermore, the local mass term is forbidden by the translational symmetry in extra spaces, 
 which is spontaneously broken by $\braket{A_{5,6}}$. 
The scalar field $\phi$ is transformed by the infinitesimal translation 
$x_{5,6}\rightarrow x_{5,6}+\epsilon_{5,6}$ and 
 we know the transformation to be a constant shift \cite{BDDS, BDD},  
\begin{align}
\delta_T \phi 
\equiv (\epsilon_5 \partial_5 + \epsilon_6 \partial_6) \phi
=\frac{\bar{\epsilon}}{\sqrt{2}}f, 
\end{align}
where $\epsilon \equiv \frac{1}{2}(\epsilon_5 + i \epsilon_6), \bar{\epsilon} \equiv \frac{1}{2}(\epsilon_5 - i \epsilon_6)$.  
This implies that $\phi$ is regarded as a NG boson of the translational symmetry in extra spaces. 
The NG boson can have only derivative couplings and the local mass term is forbidden.

For the bulk field $\Phi$, both $Z_2$ parities can be assigned 
 and $\Phi$ can be expanded in terms of mode functions $\xi_{n,j}(x^m)$ in $T^2$ flux compactification.  
\begin{align}
    \begin{array}{ll}
    \Phi(x_\mu, x_5, x_6)=\frac{1}{2L}\sum_{n=0,j}^\infty\Phi_{n,j}(x_\mu)\Bigl(\xi_{n,j}(x^m)+\xi_{n,j}(-x^m)\Bigl)& (\rm{even}), \\
    \Phi(x_\mu, x_5, x_6)=\frac{1}{2L}\sum_{n=1,j}^\infty\Phi_{n,j}(x_\mu)\Bigl(\xi_{n,j}(x^m)-\xi_{n,j}(-x^m)\Bigl)& (\rm{odd})
    \end{array}
    \label{KKexp}
\end{align}
where the summation with respect to $j$ is given by the summation for the degeneracy $\sum_{j=0}^{N-1}$.  
$\Phi_{n,j}(x_\mu)$ is a four dimensional field and 
$\xi_{n,j}(x^m), \xi_{n,j}(-x^m)$ satisfy the orthonormal conditions 
\begin{align}
\frac{1}{L^2}\int d^2x\bar{\xi}_{n',j'}(\pm x^m)\xi_{n,j}(\pm x^m)&=\delta_{n',n}\delta_{j',j},\\
\frac{1}{L^2}\int d^2x\bar{\xi}_{n',j'}(\pm x^m)\xi_{n,j}(\mp x^m)&=\delta_{-n',n}\delta_{j',j}.
\end{align}
Note that $Z_2$ even (odd) parity mode function can be constructed from the sum (subtraction)
 of mode functions $\xi_{n,j}(x^m)$ and $\xi_{n,j}(-x^m)$ in $T^2$ compactification, respectively \cite{GGH}. 

Furthermore, the mode functions of the vacuum are defined as
\begin{align}
    a\xi_{0,j}=0,\qquad a^\dagger \bar{\xi}_{0,j}=0. 
\end{align}
The mode functions with higher mode are obtained 
by acting the creation operators on the vacuum as is done in the harmonic oscillator. 
\begin{align}
    \xi_{n,j}=\frac{1}{\sqrt{n!}}(a^\dagger)^n\xi_{0,j},\qquad \bar{\xi}_{n,j}=\frac{1}{\sqrt{n!}}(a)^n\bar{\xi}_{0,j}. 
\end{align}

As for the gauge field, it can be also expanded similarly to the scalar field, 
 but the expansion does not needed since we have no gauge boson loop contributions 
 in the calculation of the quantum correction to the WL scalar mass.

We now substitute the mode expansions \eqref{KKexp} into 
the $Z_2$ even and odd terms of $\bar{\Phi}H_2\Phi$, $\bar{\Phi}a^\dagger\Phi$, 
$\bar{\Phi}a\Phi$, $\bar{\Phi}\Phi $ in \eqref{scalaraction} 
and calculate them using the properties of the creation and annihilation operators. 

First, the KK mass terms are calculated as follows. 
\begin{align}
    \bar{\Phi}H_2\Phi_{(even)}
    &=-\frac{1}{4L^2} \sum_{n=0,j} \bar{\Phi}_{n',j'}\Phi_{n,j}\Bigl\{-g_6f(2n+1) \bar{\xi}_{n'}^+\xi_n^+
    -g_6f(2n+1)\bar{\xi}_{n'}^-\xi_n^- \notag\\
    &\qquad\qquad\qquad\qquad\qquad\quad -g_6f(2n+1)\bar{\xi}_{n'}^+\xi_n^-
     -g_6f(2n+1)\bar{\xi}_{n'}^-\xi_n^+\Bigl\}, 
     \label{masseven}\\
     \bar{\Phi}H_2\Phi_{(odd)}
    &=-\frac{1}{4L^2} \sum_{n=1,j} \bar{\Phi}_{n',j'}\Phi_{n,j}\Bigl\{-g_6f(2n+1)\bar{\xi}_{n'}^+\xi_n^+
     -g_6f(2n+1)\bar{\xi}_{n'}^-\xi_n^-\notag\\
    &\qquad\qquad\qquad\qquad\qquad\quad +g_6f(2n+1)\bar{\xi}_{n'}^+\xi_n^-
    +g_6f(2n+1)\bar{\xi}_{n'}^-\xi_n^+\Bigl\}
    \label{massodd}
\end{align}
where we defined a simplified notation of the mode functions. 
\begin{align}
    \xi_{n,j}(x^m)\equiv\xi_n^+, \quad \xi_{n,j}(-x^m)\equiv\xi_n^-. 
\end{align}
Next, the terms corresponding to three point interactions are calculated. 
\begin{align}
    \bar{\Phi}a^\dagger\Phi_{(even)}
   &=\frac{1}{4L^2} \sum_{n=0,j} \bar{\Phi}_{n',j'}\Phi_{n,j}\sqrt{n+1}(\bar{\xi}_{n'}^+\xi_{n+1}^+
   +\bar{\xi}_{n'}^-\xi_{n+1}^-+\bar{\xi}_{n'}^+\xi_{n+1}^-+\bar{\xi}_{n'}^-\xi_{n+1}^+), 
   \label{3pt1}\\
     \bar{\Phi}a^\dagger\Phi_{(odd)}
   &=\frac{1}{4L^2} \sum_{n=1,j} \bar{\Phi}_{n',j'}\Phi_{n,j}\sqrt{n+1}(\bar{\xi}_{n'}^+\xi_{n+1}^+
   +\bar{\xi}_{n'}^-\xi_{n+1}^--\bar{\xi}_{n'}^+\xi_{n+1}^--\bar{\xi}_{n'}^-\xi_{n+1}^+), 
   \label{3pt2}\\
    \bar{\Phi}a\Phi_{(even)}
    &=\frac{1}{4L^2} \sum_{n=0,j} \bar{\Phi}_{n',j'} \Phi_{n,j} \sqrt{n}(\bar{\xi}_{n'}^+\xi_{n-1}^+
    +\bar{\xi}_{n'}^-\xi_{n-1}^-+\bar{\xi}_{n'}^+\xi_{n-1}^-+\bar{\xi}_{n'}^-\xi_{n-1}^+), 
    \label{3pt3}\\
    \bar{\Phi}a\Phi_{(odd)}
    &=\frac{1}{4L^2} \sum_{n=1,j} \bar{\Phi}_{n',j'} \Phi_{n,j} \sqrt{n}(\bar{\xi}_{n'}^+\xi_{n-1}^+ 
    +\bar{\xi}_{n'}^-\xi_{n-1}^--\bar{\xi}_{n'}^+\xi_{n-1}^--\bar{\xi}_{n'}^-\xi_{n-1}^+). 
    \label{3pt4}
\end{align}
Finally, a part of the quadratic interaction are calculated. 
\begin{align}
    \bar{\Phi}\Phi_{(even)} 
    &=\frac{1}{4L^2} \sum_{n=0,j} \bar{\Phi}_{n',j'} \Phi_{n,j} 
    (\bar{\xi}_{n'}^+\xi_n^++\bar{\xi}_{n'}^-\xi_n^-+\bar{\xi}_{n'}^+\xi_n^-+\bar{\xi}_{n'}^-\xi_n^+), 
    \label{4pt1}\\
    \bar{\Phi}\Phi_{(odd)}
    &= \frac{1}{4L^2} \sum_{n=1,j} \bar{\Phi}_{n',j'} \Phi_{n,j}
    (\bar{\xi}_{n'}^+\xi_n^++\bar{\xi}_{n'}^-\xi_n^--\bar{\xi}_{n'}^+\xi_n^--\bar{\xi}_{n'}^-\xi_n^+). 
\label{4pt2}
\end{align}

\section{One-loop correction to WL scalar mass}
In this section, we calculate one-loop corrections to the WL scalar mass 
 by the KK modes of the scalar field $\Phi$. 

Using these results \eqref{masseven}-\eqref{massodd}, \eqref{3pt1}-\eqref{3pt4}, \eqref{4pt1}-\eqref{4pt2} 
 and the orthonormal conditions of the mode functions, 
 we can derive the four dimensional effective theory. 
Since the action with nonzero modes is obtained from the sum of the $Z_2$ even and odd terms, 
 but the action with only zero mode is obtained from the $Z_2$ even terms,  
 we summarize the four dimensional effective actions for nonzero KK modes and zero mode of $\Phi$, separately. 
\begin{align}
    S_{4,n\ne0} &= \int d^4x \biggl[
    -\frac{1}{4} \sum_{n=1}^{\infty}F_{n}^{\mu\nu} F_{n\mu\nu} 
    - \partial_{\mu} \overline{\varphi} \partial^{\mu} \varphi \notag\\    
    &\qquad\qquad
    + \sum_{n=1,j}^\infty \Bigl(
    -(D_{\mu}\Phi_{n,j})^* D^{\mu} \Phi_{n,j} -2g_4 fL \Bigl( n+\frac{1}{2} \Bigl) \bar{\Phi}_{n,j} \Phi_{n,j} 
    \notag\\
    &\qquad\qquad -\sqrt{2} ig_4 \sqrt{2g_4fL(n+1)} \overline{\varphi} \bar{\Phi}_{n+1,j}\Phi_{n,j} 
    +\sqrt{2}ig_4 \sqrt{2g_4fL(n+1)} \varphi\bar{\Phi}_{n,j} \Phi_{n+1,j} \notag\\
     &\qquad\qquad\qquad -2g_4^2\overline{\varphi}\varphi\bar{\Phi}_{n,j}\Phi_{n,j}
     \Bigl)\biggl]
     \label{4Deff1}
\end{align}
for non-zero modes and 
\begin{align}
    S_{4,n=0} &= \int d^4x \biggl[
    -\frac{1}{4}F_0^{\mu\nu}F_{0\mu\nu} - \partial_{\mu} \overline{\varphi}\partial^{\mu} \varphi \notag\\
    &\qquad\qquad +\Bigl(
    -(D_{\mu}\Phi_{0,j})^* D^{\mu}\Phi_{0,j} -g_4fL \bar{\Phi}_{0,j} \Phi_{0,j} \notag\\
    &\qquad\qquad\qquad 
    -\frac{\sqrt{2}}{2}ig_4 \sqrt{2g_4fL} \overline{\varphi} \bar{\Phi}_{1,j} \Phi_{0,j} 
    -\frac{\sqrt{2}}{2}ig_4 \sqrt{2g_4fL} \overline{\varphi}\bar{\Phi}_{-1,j} \Phi_{0,j} \notag\\
    &\qquad\qquad\qquad 
    +\frac{\sqrt{2}}{2}ig_4 \sqrt{2g_4fL} \varphi\bar{\Phi}_{0,j} \Phi_{1,j} 
    +\frac{\sqrt{2}}{2}ig_4 \sqrt{2g_4fL} \varphi\bar{\Phi}_{0,j} \Phi_{1,j} \notag\\
    &\qquad\qquad\qquad -2g_4^2\overline{\varphi}\varphi\bar{\Phi}_{0,j}\Phi_{0,j}
    \Bigl)\biggl]
    \label{4Deff2}
\end{align}
for zero mode. 
The four dimensional gauge coupling is defined as $Lg_4=g_6$. 
We note that the mixing terms $\bar{\xi}^+\xi^-$,$\bar{\xi}^-\xi^+$ are cancelled in calculation of \eqref{4Deff1}. 

In order to calculate the quantum corrections to the mass of $\varphi$, 
 we first derive the propagator of the scalar field, which is denoted as $K^{-1}(p)$ 
 in a four dimensional point of view \cite{GGH}. 
The free part Lagrangian of the scalar field in the four dimensional effective theory are written as    
\begin{align}
{\cal L} &=-\frac{1}{2} \sum_{n,j} \bar{\Phi}_{n',j'} 
    \biggl[(\delta_{n',n}\pm \delta_{-n',n})\Bigl\{-\partial^2 + 2g_4fL\Bigl(n+\frac{1}{2}\Bigl)\Bigl\}\biggl] \Phi_{n,j},
\end{align}
then the propagator can be derived as
\begin{align}
    K^{-1}(p)&=\frac{-i}{2}\Biggl(\frac{\delta_{n',n} \pm \delta_{-n',n}}{p^2+2g_4fL\Bigl(n+\frac{1}{2}\Bigl)}\Biggl). 
\label{propagator}
\end{align}
$+(-)$ corresponds to the $Z_2$ even (odd) case, respectively.   
It is straightforward to obtain the propagator of  nonzero modes $\Phi_{n,j}~(n\ne0)$ 
from \eqref{propagator} by adding the both propagators of $Z_2$ even and odd cases 
and the result is
\begin{align}
    K^{-1}(p)=\frac{-i\delta_{n',n}}{p^2+2g_4fL\Bigl(n+\frac{1}{2}\Bigl)}. 
    \label{eq:500}
\end{align}

For zero mode, we should consider the $Z_2$ even and odd part separately.  
In \eqref{propagator}, $n,n'=0$ is substituted and the propagator of the zero mode is obtained. 
\begin{align}
  K^{-1}(p) &\to \frac{-i}{p^2+g_4fL}~({\rm even}), \quad 0~({\rm odd}). 
\end{align}
It is consistent that the propagator of zero mode is obtained from the even part. 
We emphasize that the $\delta_{-n',n}$ term in the propagator is characteristic in the case of $T^2/Z_2$, 
 and is absent in the $T^2$ compactification. 
This term reflects the violation of KK momentum conservation on the fixed points \cite{GGH} 
 and plays an important role in calculation of the quantum correction to the mass of $\varphi$. 

\begin{figure}[h]
    \centering
    \includegraphics[width=40mm]{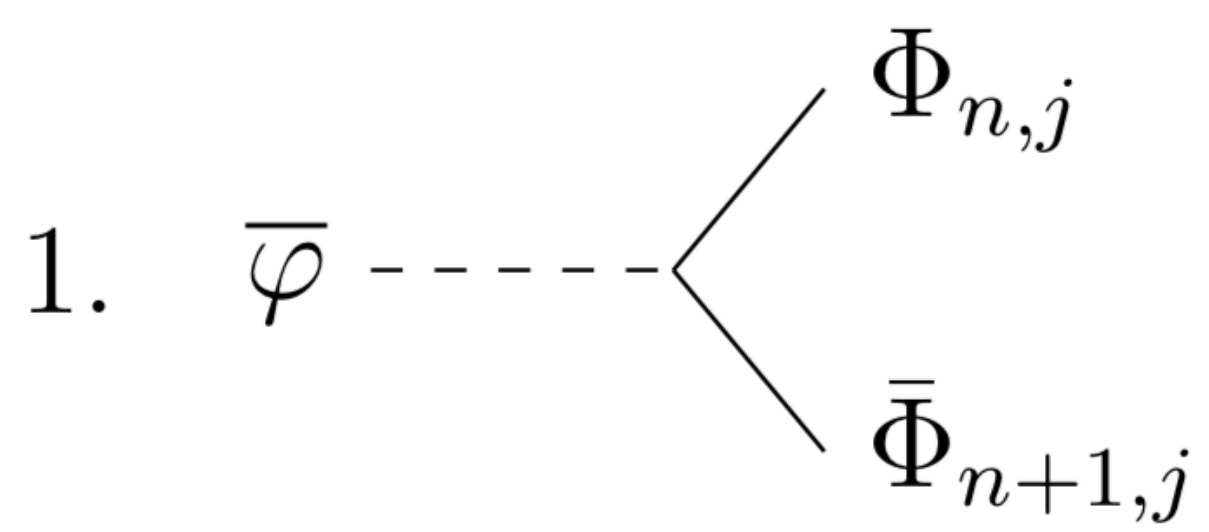}
    \quad
 \includegraphics[width=40mm]{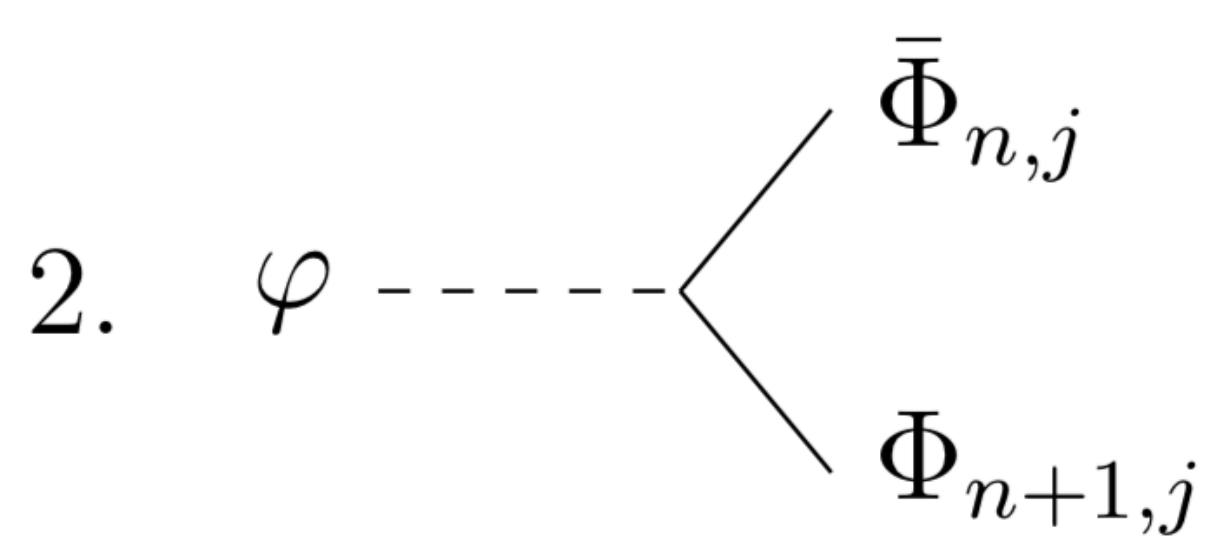}
 \quad
  \includegraphics[width=42mm]{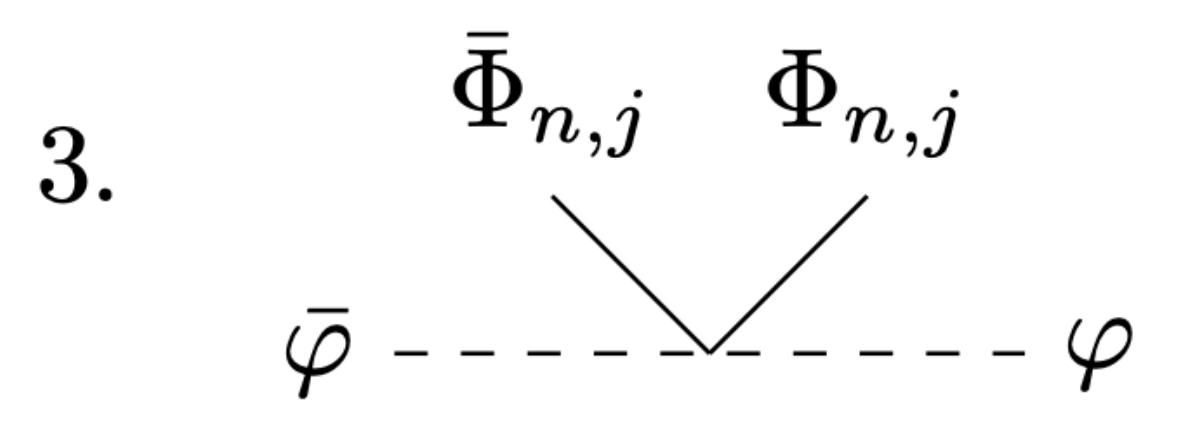}
      \caption{3-point and 4-point vertices from \eqref{4Deff1}, 
    which include only non-zero modes for the bulk scalar field.}
    \label{fig:nneq00}
\end{figure}

\begin{figure}[h]
    \centering
    \includegraphics[width=38mm]{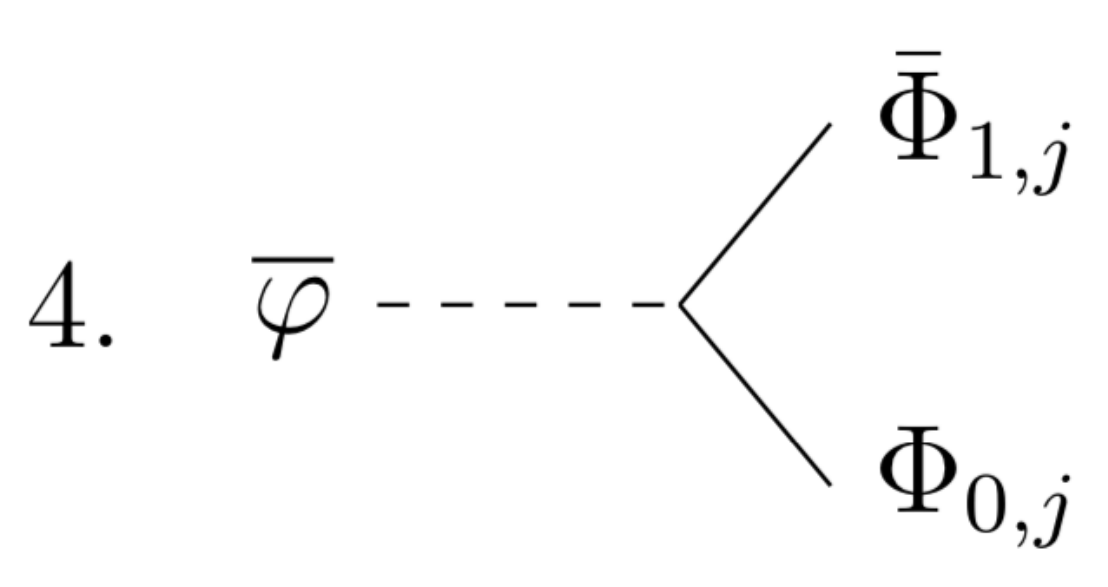}
    \quad
 \includegraphics[width=38mm]{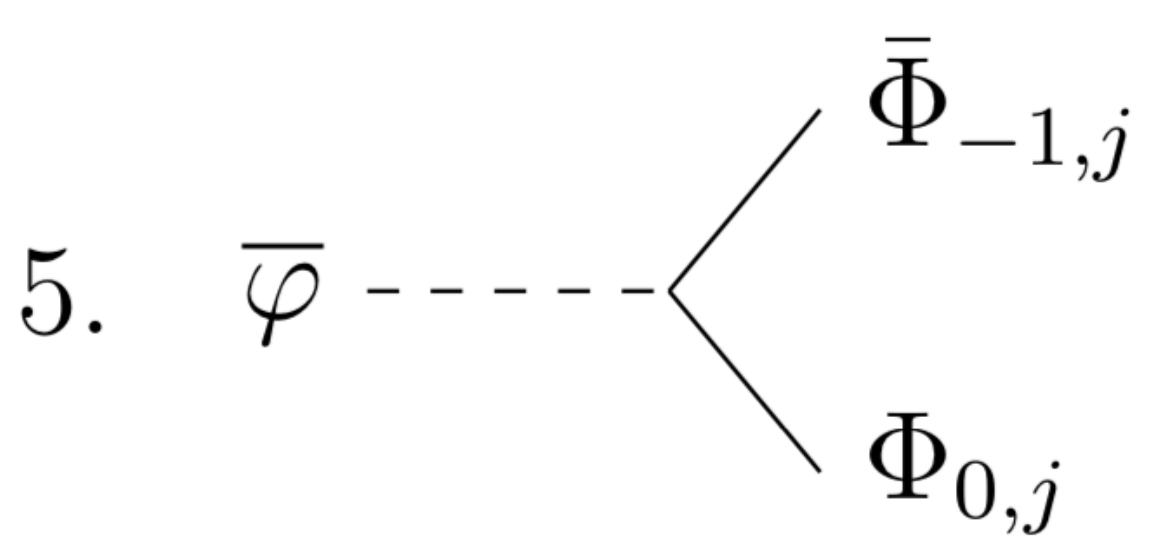}
 \quad
  \includegraphics[width=38mm]{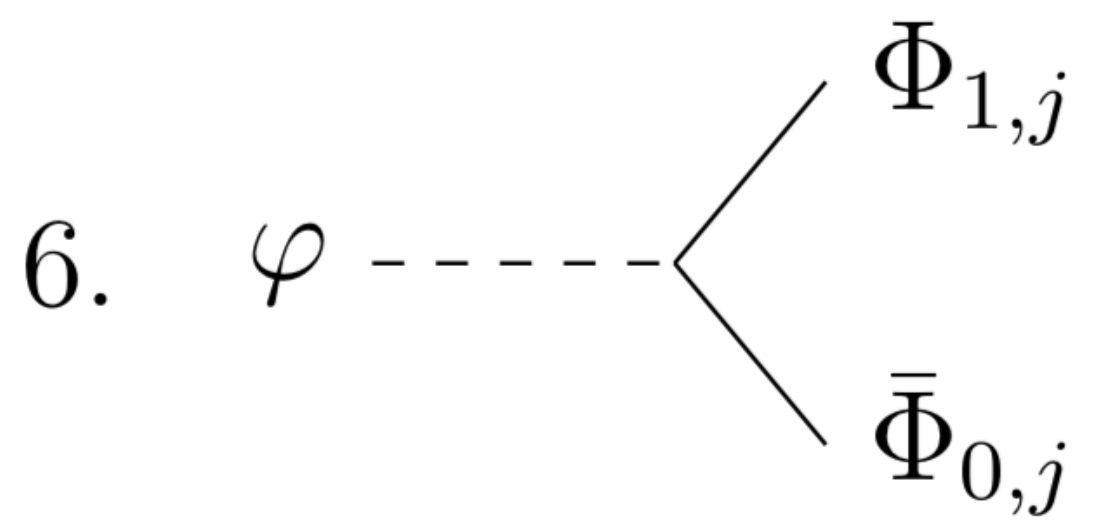} \\
  \vspace*{5mm}
    \includegraphics[width=38mm]{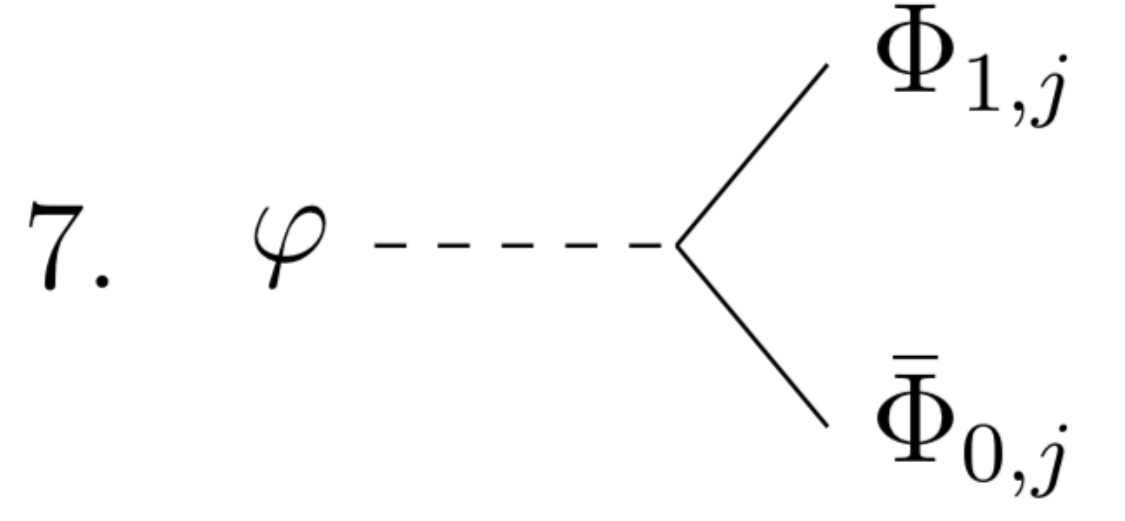}
    \quad
 \includegraphics[width=42mm]{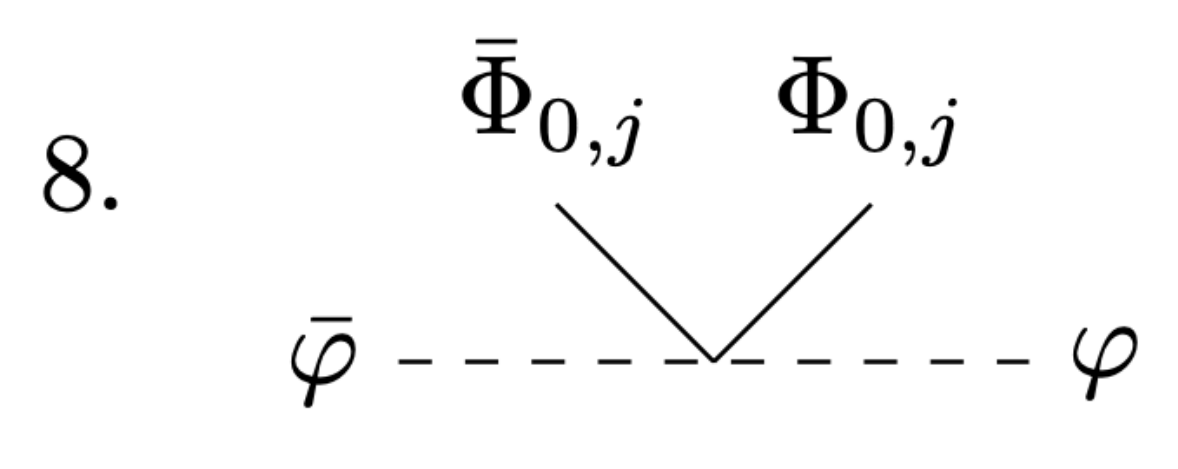}
    \caption{3-point and 4-point vertices from \eqref{4Deff2}, 
    which include at least one zero mode for the bulk scalar field.}
    \label{fig:n=0fin}
\end{figure}

The interaction vertices necessary for one-loop calculations are derived
 and they are shown in Figure \ref{fig:nneq00} and Figure \ref{fig:n=0fin}. 
We note that the 3-point vertices 6 and 7 in Figure \ref{fig:n=0fin} are apparently the same, 
 but are coming from different interaction terms.  
The relevant 3-point interactions in parity even Lagrangian are
\begin{align}
    & -\frac{1}{2}\sqrt{2} ig_4 \sqrt{2g_4fL(n+1)} \bar{\varphi} \bar{\Phi}_{n+1,j} \Phi_{n,j}
    -\frac{1}{2} \sqrt{2} ig_4 \sqrt{2g_4fL(n+1)} \bar{\varphi} \bar{\Phi}_{-n-1,j} \Phi_{n,j}\notag\\
    &+\frac{1}{2} \sqrt{2}ig_4 \sqrt{2g_4fL(n+1)} \varphi\bar{\Phi}_{n,j}\Phi_{n+1,j}
    + \frac{1}{2} \sqrt{2}ig_4 \sqrt{2g_4fL(n+1)}\varphi\bar{\Phi}_{-n,j} \Phi_{n+1,j}. 
    \end{align} 
The 3-point vertices 6 and 7 are obtained from the third and the fourth terms in the above Lagrangian 
 after putting $n=0$, respectively.

Let us first calculate one-loop contributions to WL scalar mass from $n\ne0$ KK modes.   
The one-loop corrections by the two 3-point vertices (1 and 2 in Figure \ref{fig:nneq00}) is calculated as
\begin{align}	
\delta m^2_\varphi (3{\rm -point})
&=-i (\sqrt{2}g_4)^2 \sum_{n=1,j}^{\infty}\int \frac{d^4k}{(2\pi)^4}
 \Biggl(\frac{-i\sqrt{2g_4fL(n+1)}}{k^2+2g_4fL\bigl(n+\frac{1}{2}\bigl)}\Biggl)
 \Biggl(\frac{-i\sqrt{2g_4fL(n+1)}}{k^2+2g_4fL\bigl(n+\frac{3}{2}\bigl)}\Biggl)\notag\\
        &=i2g_4^2 |N| \sum_{n=1}^{\infty}\int \frac{d^4k}{(2\pi)^4} 
        \frac{2g_4fL(n+1)}{\Bigl(k^2+2g_4fL\bigl(n+\frac{1}{2}\bigl)\Bigl)\Bigl(k^2+2g_4fL\bigl(n+\frac{3}{2}\bigl)\Bigl)}. 
        \label{nonzero3pt}
\end{align}
Similarly, the one-loop corrections by a 4-point vertex (3 in Figure \ref{fig:nneq00}) is obtained as
\begin{align}
\delta m^2_\varphi(4{\rm -point})
= 2g_4^2 |N| \sum_{n=1}^{\infty}\int \frac{d^4k}{(2\pi)^4} \frac{1}{k^2+2g_4fL\Bigl(n+\frac{1}{2}\Bigl)}. 
\label{nonzero4pt}
\end{align}
In the final expressions of (\ref{nonzero3pt}) and (\ref{nonzero4pt}), Wick rotation has been done.

Thus, the one-loop corrections to the WL scalar mass by the nonzero KK modes of the scalar field results in
 \begin{align}
        \delta m_\varphi^2(n\ne0) &= 2g_4^2 
        \sum_{n=1,j}^{\infty} \int \frac{d^4k}{(2\pi)^4} \notag\\
  &\times \Biggl\{\frac{1}{k^2+2g_4fL(n+\frac{1}{2})} - \frac{2g_4fL(n+1)}{\Bigl(k^2+2g_4fL\bigl(n+\frac{1}{2}\bigl)\Bigl)
  \Bigl( k^2 + 2g_4fL \bigl( n+\frac{3}{2} \bigl) \Bigl)} 
  \Biggl\}\notag\\
        &=2g_4^2|N| \sum_{n=1}^{\infty}\int \frac{d^4k}{(2\pi)^4} 
        \Biggl(\frac{n+1}{k^2+2g_4fL\bigl(n+\frac{3}{2}\bigl)}-\frac{n}{k^2+2g_4fL(n+\frac{1}{2})}\Biggl). 
        \label{eq:nnot0}
    \end{align}
    
Next, we turn to consider zero mode $n=0$ contributions. 
The one-loop corrections by the two 3-point vertices are calculated 
 by connecting the vertices 4 and 6 in Figure \ref{fig:n=0fin} and the vertices 5 and 7 in Figure \ref{fig:n=0fin}, respectively. 
We note that the former contributions are connected by the $\delta_{n', n}$ term in the propagator, 
 but the latter contributions are connected by the $\delta_{-n', n}$ term in the propagator, 
 which are absent in the $T^2$ compactification. 
These contributions are found to be same and the result is  
\begin{align}
    \delta m^2_\varphi(3{\rm -point}) 
    &= -i \Biggl(\frac{\sqrt{2}}{2}g_4\Biggl)^2 
    \sum_{j}\int \frac{d^4k}{(2\pi)^4} \Biggl(\frac{-i\sqrt{2g_4fL}}{k^2+g_4fL}\Biggl) 
    \Biggl(\frac{-i\sqrt{2g_4fL}}{k^2+3g_4fL}\Biggl) \times 2 \notag\\
    &=-g_4^2|N| \int \frac{d^4k}{(2\pi)^4} 
    \frac{2g_4fL}{\Big(k^2+g_4fL\Bigl)\Big(k^2+3g_4fL\Bigl)}. 
    \label{eq:3situ}
\end{align}
The one-loop correction by a 4-point vertex (8 in Figure \ref{fig:n=0fin}) is easily calculated 
\begin{align}
    \delta m^2_\varphi(4 {\rm -point})
     = 2g_4^2|N| \int \frac{d^4k}{(2\pi)^4} \frac{1}{k^2+g_4fL}. 
    \label{eq:4situ}
\end{align}
In the final expression of \eqref{eq:3situ} and \eqref{eq:4situ}, Wick rotation has been done. 

Therefore, one-loop corrections to the WL scalar mass from zero mode contributions 
 is a sum of \eqref{eq:3situ} and \eqref{eq:4situ}
\begin{align}
\delta m^2_\varphi(n=0)
&=-g_4^2|N| \int \frac{d^4k}{(2\pi)^4} \frac{2g_4fL}{\Big(k^2+g_4fL\Bigl)\Big(k^2+3g_4fL\Bigl)}
+2g_4^2|N| \int \frac{d^4k}{(2\pi)^4} \frac{1}{k^2+g_4fL}\notag\\
&=g_4^2|N|\int \frac{d^4k}{(2\pi)^4} \Bigl(\frac{1}{k^2+3g_4fL}+\frac{1}{k^2+g_4fL}\Bigl). 
\label{eq:n=0s}
\end{align}
Finally, we obtain one-loop corrections to the WL scalar mass 
 by summing the results \eqref{eq:nnot0} and \eqref{eq:n=0s}. 
\begin{align}
\delta m_\varphi^2
&=2g_4^2|N| \sum_{n=1}^{\infty}\int \frac{d^4k}{(2\pi)^4} 
\Biggl(\frac{n+1}{k^2+2g_4fL\bigl(n+\frac{3}{2}\bigl)}-\frac{n}{k^2+2g_4fL(n+\frac{1}{2})}\Biggl)\notag\\
&\quad+g_4^2|N|\int \frac{d^4k}{(2\pi)^4} \Bigl(\frac{1}{k^2+3g_4fL}+\frac{1}{k^2+g_4fL}\Bigl)\notag\\
&=2g_4^2|N| \sum_{n=0}^{\infty}\int \frac{d^4k}{(2\pi)^4} 
\Biggl(\frac{n+1}{k^2+2g_4fL\bigl(n+\frac{3}{2}\bigl)}-\frac{n}{k^2+2g_4fL(n+\frac{1}{2})}\Biggl)\notag\\
&\quad+g_4^2|N|\int \frac{d^4k}{(2\pi)^4} \Bigl(-\frac{1}{k^2+3g_4fL}+\frac{1}{k^2+g_4fL}\Bigl)\notag\\
&=g_4^2|N|\int\frac{d^4k}{(2\pi)^4} \Bigl(\frac{1}{k^2+g_4fL}-\frac{1}{k^2+3g_4fL}\Bigl) 
\label{eq:result}\\ 
&\to g_4^2|N|\frac{g_4fL}{16\pi^2}\log\Bigl(\frac{\Lambda^4}{27(g_4fL)^2}\Bigl). 
\end{align}
Noting that the third term in the right-hand side of the first equality is 
a half of the first term for $n=0$ case, 
we rewrite the mode summation from $n=1$ to that from $n=0$ in the second equality. 
Furthermore, the rewritten mode summation vanishes by the shift $n \to n+1$ in the second term, 
 which is ensured from the translational symmetry in extra spaces of $T^2$ compactification.  
As expected, the final result (\ref{eq:result}) is expressed by only zero mode 
and is interpreted that the mass of the WL scalar field is generated at fixed points.  
Remarkably, the degree of divergence is found to be logarithmic not quadratic, 
 which is very similar to supersymmetric case. 
This is considered to be a remnant of translational symmetry in extra spaces. 
In the final expression, the momentum integral is explicitly done by introducing the cutoff scale $\Lambda$.  


One might think a following question. 
It is natural that the obtained WL scalar mass is generated at the fixed points, 
 where the translational symmetry in extra spaces is explicitly broken. 
However, it might be inconsistent that the WL scalar mass is generated on the fixed points 
 since the gauge symmetry is operative on the fixed points as mentioned before. 
To resolve this question, we have to take care of the meaning of mass. 
The WL scalar mass obtained in this paper should be understood 
 as a gauge invariant and non-local mass described by the Wilson-line operators \cite{HM3}. 
The Wilson-line operators are given by 
\begin{align}
U_5=\exp\Bigl[ig_6\oint A_5dx^5\Bigl],\qquad U_6=\exp\Bigl[ig_6\oint A_6dx^6\Bigl],
\end{align}
which can be rewritten as follows
\begin{align}
U_5 &= \exp \Bigl[ \frac{g_4}{\sqrt{2}} \oint (\varphi dz + \varphi d\bar{z} 
 -\bar\varphi dz - \bar\varphi d\bar{z})\Bigl], \\
U_6 &= \exp \Bigl[ \frac{g_4}{\sqrt{2}} \oint (\varphi dz - \varphi d\bar{z} 
 +\bar\varphi dz - \bar\varphi d\bar{z})\Bigl]. 
 \label{WLop}
\end{align}
From this observation, the WL scalar field can be expressed 
 by the gauge invariant non-local operator as 
\begin{align}
\varphi \propto U_5-U_5^\dagger, \qquad
\bar\varphi \propto U_6-U_6^\dagger
\end{align}
and the obtained WL scalar mass is gauge invariant and non-local.  
Furthermore, we can verify that these Wilson-line operators $U_{5,6}$ are not invariant 
 under the constant shift. 

\section{Conclusions}
In this paper, we have calculated one-loop quantum corrections to the WL scalar mass 
 in a six dimensional scalar QED compactified on an orbifold $T^2/Z_2$ with magnetic flux. 
In $T^2$ compactification with magnetic flux, the WL scalar field is 
 the massless NG boson of the translational symmetry in extra spaces. 
In a realistic situation, we would like to regard the WL scalar field as the SM Higgs scalar field 
 in a spirit of gauge-Higgs unification.   
However, this is impossible as it stands because of the masslessness of the WL scalar field. 
In order to generate the mass, we have to break the translational symmetry in extra spaces explicitly 
 and the NG boson becames a pseudo NG boson. 
In this paper, by employing the $T^2/Z_2$ orbifold compactification as one of the solutions, 
 the symmetry is explicitly broken on fixed points in $T^2/Z_2$ space. 
Then, the nonvanishing scalar mass is expected to be generated on fixed points.  
In fact, we have shown that this is true by calculating one-loop corrections to the WL scalar mass 
 in a six dimensional scalar QED compactified on an orbifold $T^2/Z_2$ with magnetic flux. 
The contributions of the corrections to the WL scalar mass were found to be from only the zero mode. 
This is a natural result since only the four dimensional spacetime is spread on the fixed points.  
Remarkably, the nonvanishing scalar mass is logarithmic divergent not quadratic divergent, 
 which is similar to the supersymmetric theory. 
 
From the viewpoint of constructing a realistic model, 
 it is necessary to extend the present discussion to the non-Abelian theory. 
In this extension, there exist extra contributions of the quantum corrections to the WL scalar mass 
 from the gauge bosons, the scalar fields from the extra component of the gauge field, and the ghost fields. 
It would be interesting if the logarithmic divergences generated in the WL scalar mass are canceled to be finite 
 and the cutoff scale independent by combining the above contributions.  
If this is true, it would be more desirable from the viewpoint of solving the hierarchy problem. 
This issue is remained for future work. 

\section*{Acknowledgments}
We thank Takuya Hirose for discussion. 



\begin{thebibliography}{99}


\bibitem{BDDS}
W.~Buchmuller, M.~Dierigl, E.~Dudas and J.~Schweizer,
JHEP \textbf{04}, 052 (2017) [arXiv:1611.03798 [hep-th]].

\bibitem{BDD}
W.~Buchmuller, M.~Dierigl and E.~Dudas,
JHEP \textbf{08}, 151 (2018) [arXiv:1804.07497 [hep-th]]. 

\bibitem{GL}
D.~M.~Ghilencea and H.~M.~Lee,
JHEP \textbf{06}, 039 (2017)
[arXiv:1703.10418 [hep-th]].

\bibitem{HM1}
T.~Hirose and N.~Maru,
JHEP \textbf{08}, 054 (2019) [arXiv:1904.06028 [hep-th]].

\bibitem{HM2}
T.~Hirose and N.~Maru,
J. Phys. G \textbf{48}, no.5, 055005 (2021) [arXiv:2012.03494 [hep-th]].

\bibitem{HS}
M.~Honda and T.~Shibasaki,
JHEP \textbf{03}, 031 (2020) [arXiv:1912.04581 [hep-th]].


\bibitem{HM3}
T.~Hirose and N.~Maru,
JHEP \textbf{06}, 159 (2021) [arXiv:2104.01779 [hep-th]].

\bibitem{GIQ}
G.~von Gersdorff, N.~Irges and M.~Quiros,
Nucl. Phys. B \textbf{635}, 127-157 (2002)
[arXiv:hep-th/0204223 [hep-th]].

\bibitem{GGH}
H.~Georgi, A.~K.~Grant and G.~Hailu,
Phys. Lett. B \textbf{506}, 207-214 (2001) [arXiv:hep-ph/0012379 [hep-ph]].




\end{thebibliography}
\end{document}